# Hot tori around black holes as sources of gamma ray bursts

M. Jaroszyński[1,2]

[1] Warsaw University Observatory, Al. Ujazdowskie 4, 00–478 Warsaw, Poland
[2] D.A.R.C., Observatoire Paris-Meudon, 5 Pl. Janssen, 92195 Meudon Cedex, France

**Abstract.** We investigate the configurations consisting of massive, dense and hot tori around stellar mass Kerr black holes as possible sources of energy for the gamma ray bursts in cosmological hypothesis of their origin. We limit parameters of our models to the values resulting from neutron stars merger calculations or suggested in the "failed supernova" scenario. We investigate models with different angular momentum distributions and different specific entropies. We construct also approximate evolutionary tracks of our systems postulating some viscosity mechanism to be present. We find, that models resulting from a merger of two neutron stars give too little energy to be likely sources of gamma ray bursts. This conclusion remains true despite the artificially high values of specific entropy and viscosity, which we use in our calculations. The promising models should contain a high angular momentum black hole ($a \sim 1$) and/or a torus with almost constant specific angular momentum. The configurations resulting from a collapse of rapidly rotating WR stars are not excluded as sources of the bursts due to a greater freedom in choosing their initial parameters.

## 1. Introduction

The nature of classical gamma ray bursts is still unknown. The isotropy of the bursts on the sky (Meegan et al. 1992) and the deficit of the weak (more distant) events (Fenimore et al. 1993) makes the cosmological hypothesis of their origin likely.

The nonthermal spectrum of bursts, extending up to GeV range in some cases (Sommer et al. 1994) is another problem. According to Mészáros & Rees (1993) a nonthermal spectrum can be produced if a relativistically expanding *fireball* interacts with circumstellar material. The fireball should be "clean" (Shemi & Piran 1990) i.e. containing low fraction of baryons. The annihilation of neutrinos close to the axis of a rotating system, where the density of baryons should be low due to the centrifugal force barrier, has been proposed as a way to produce a clean fireball (Eichler et al. 1989, Mészáros & Rees 1992).

For the sources distributed cosmologically, up to a redshift $z \sim 1$ their total energy should be of the order of $10^{51}$ ergs - similar to energies of a supernova (Paczyński 1986). The variability time scales, $< 1$ ms, are similar to dynamical time scale on the surface of a neutron star (Narayan et al., 1992). These coincidences suggests, that the bursts originate near neutron stars. Several mechanisms of this kind are considered by Narayan et al. (1992). In this paper we limit ourselves to scenarios involving the coalescence of two neutron stars or a collapse of rapidly rotating (Bodenheimer & Woosley 1983) Wolf-Rayet star (Woosley 1993), producing a black hole surrounded with a massive disk. This is the continuation of the approach of Mochkovitch et al. (1993), Jaroszyński (1993) and Witt et al. (1994).

Our model serves as a primary energy source of the fireball. The temporal characteristic of the burst radiation, which is emitted later in the shocks, can be very different from the "light" curves produced by our calculations. The total energy delivered to the fireball must be at least equal to the energy required by bursts observations (assuming 100% efficiency of converting kinetic energy into gamma ray radiation). Fenimore et al. (1993) fit a "standard candle" model to the observed flux - number relation for the bursts and obtain a typical source luminosity $L \sim 6 \times 10^{50}$ erg s$^{-1}$. For a typical bursts duration of $\sim 10$ s the total emitted energy should be greater than $\sim 10^{51}$ ergs . In the case of a source emitting beams of energy into a solid angle $\delta\Omega$ the above estimates should be corrected by a factor $\delta\Omega/4\pi$; still the full sky values can be used as parameters, which are easy to compare between different models, and we follow this convention. We check whether models of investigated class are capable of producing the required amount of energy in the form of relativistically expanding plasma jet. We try to make our models realistic, limiting from the begining their possible parameter values to the set suggested by neutron star

1994, Zhuge et al. 1994) or by the *failed supernova* scenario (Woosley 1993).

In the next section we describe our model with some details. In Sec.3 we describe the way of adopting model parameters. Section 4 shows our results for tori with or without effective viscosity mechanism. The discussion and conclusions follow.

## 2. Theoretical model

### 2.1. Torus structure

Following Witt et al. (1994) and Jaroszyński (1993) we use the barotropic, test fluid configurations in the external field of a Kerr black hole as models of neutrino sources. Since the specific angular momentum of matter is not constant in configurations resulting from calculated mergers of binary neutron star systems (see below) we consider a more general case of power law distribution of specific angular momentum and angular velocity. In the case of barotropic equation of state, the structure of the disk is defined by the angular momentum distribution (Abramowicz et al. 1978, hereafter AJS). In Newtonian mechanics a power law distribution of angular velocity implies a power law distribution of specific angular momentum:

$$\Omega_N \sim R^{-\beta} \quad l_N \sim R^{2-\beta} \qquad (1)$$

where $R$ is the cylindrical coordinate measuring the distance from the orbital axis and $\beta$ is a constant. Eliminating $R$ from these relations we get:

$$l_N \sim \Omega_N^{-(2-\beta)/\beta} \qquad (2)$$

Relativistic definitions of angular velocity $\Omega$ and momentum $l$ are different (see AJS), but asymptotically, for large radii, they agree with the Newtonian definitions. The power law distribution of angular momentum in the frame of general relativity can be defined as the relation between relativistic quantities in the form of the above equation. Following AJS we introduce the effective potential $W$ and using their results we obtain:

$$W = \ln(u_t) - \frac{2-\beta}{2\beta-2} \ln(1-\Omega l) \qquad (3)$$

where $u_t$ is the energy of a unit mass particle moving with the fluid four-velocity. The effective potential vanishes at infinity. The angular momentum at any point is found iteratively from the relation:

$$l \, g_{tt} + (1+\Omega l) \, g_{t\phi} + \Omega \, g_{\phi\phi} = 0 \qquad (4)$$

which is a consequence of angular momentum and velocity definitions in General Relativity (see i.e. AJS for more details). $g_{ij}$ are the metric components in the standard Boyer-Lindquist coordinate system measured at the point case. For a barotropic equation of state the hydrostatic equilibrium equation can be integrated, to give (AJS):

$$\int_0^P \frac{dP}{\epsilon+P} = W_{surf} - W \qquad (5)$$

where $P$ is the pressure $\epsilon$ the energy density and $W$ the effective potential value at a given point; $W_{surf}$ is the potential value on the fluid surface. For a barotropic equation of state all thermodynamic parameters are functions of one parameter only. Reversing the above relation we can make them functions of the effective potential (i.e. $P = P(W_{surf} - W)$, $\epsilon = \epsilon(W_{surf} - W)$ etc).

Various global characteristics of the torus can be obtained as integrals over the configuration. The baryonic mass is given as:

$$M_{torus} = \int \rho_0 u^t \sqrt{-g} dr d\theta d\phi \qquad (6)$$

where $\rho_0$ is the rest mass density of matter, $g$ is the metric determinant and $r$, $\theta$, $\phi$ are Boyer-Lindquist spherical coordinates. For the total energy of the torus treated as a test fluid we have (Bardeen, 1973b):

$$E_{torus} = \int [(\epsilon+P)u^t u_t - P] \sqrt{-g} dr d\theta d\phi \qquad (7)$$

and for the angular momentum (ibid.):

$$J_{torus} = \int (\epsilon+P) u^t u_\phi \sqrt{-g} dr d\theta d\phi \qquad (8)$$

To find the internal energy in the torus we compare the energy of the actual configuration with a thin Keplerian disk consisting of cold matter with the same distribution of rest mass with angular momentum:

$$E_{torus}^{int} = E_{torus} - \int u_t^{Kep}(l) \, \frac{dM_{torus}}{dl} \, dl \qquad (9)$$

where $u_t^{Kep}$ is the energy of a unit mass particle on a stable Keplerian orbit with given specific angular momentum. In principle even a cold disk may not be completely thin due to the non vanishing pressure of degenerate matter. We neglect this effect and treat $E_{torus}^{int}$ as energy, that could be radiated away from the torus if any viscosity mechanism were absent. In the presence of viscosity the angular momentum distribution can be changed releasing some extra energy. The effectiveness of such process depends on the nature of viscosity. We shall use a phenomenological approach, so called $\alpha$-disk theory (Shakura & Sunyaev 1973) to estimate the rate of viscous heat generation in the torus. In this approach one assumes that the viscous stress component acting accros the cylinder of constant angular velocity is proportional to the pressure, $\alpha$ being the coefficient of proportionality. For the power law angular velocity distribution the value of shear is given as

rate of viscous heating can be estimated as:

$$Q^+_{visc} \approx \alpha \int \beta \Omega P \sqrt{-g} dr d\theta d\phi. \qquad (10)$$

## 2.2. Equation of state

We assume the specific entropy of matter $S$ to be constant in space for any of the considered configurations. This is mostly a technical simplification but it may also be a physical reality in some cases (see the discussion). We consider configurations with different values of $S$ (entropy of all matter constituents in Boltzmann constant units per baryon). In this paper we consider rather hot tori with $S \geq 4$. (Witt et al. 1994 and Jaroszyński 1993 use $S = 3$, but their tori are too cold to produce a powerful fireball).

We are interested in configurations with trapped neutrinos and we use lepton number $Y_L \equiv (n_e - n_{\bar{e}} + n_\nu - n_{\bar{\nu}})/n_B$ including electrons and three neutrino flavors. Its value depends on the composition of matter in the torus. Tori being a result of neutron stars merger should have a low lepton fraction ($Y_L = 0.05$ - $0.30$, depending on the equation of state, Weber & Weigel 1989). We adopt $Y_L = 0.20$ as a representative value. In the case of a torus made during a Wolf-Rayet star collapse the matter consists of heavy nuclei and the lepton fraction is $Y_L \approx 0.45$.

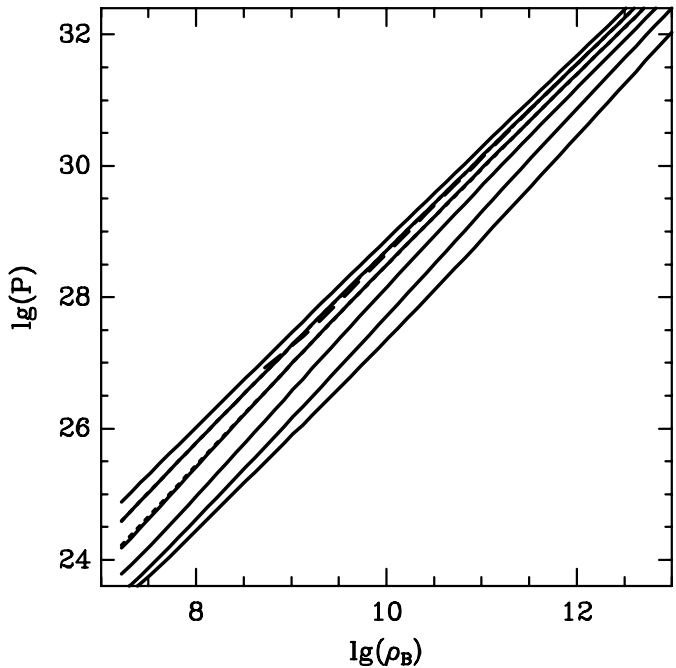

**Fig. 1.** Pressure versus rest mass density for different equations of state. Our approximate EOS for $Y_L = 0.2$ $S = 4, 6, 8, ..., 14$ is drawn using solid lines. The dotted lines show our results for $Y_L = 0.4$ and $S = 10, 12$, while the dashed line shows the result of the Lattimer & Swesty equation of state with $Y_L = 0.4$ and $S = 11$, for comparison.

proximate equation of state. We assume the matter consists of free protons, neutrons, electrons, positrons, three neutrino flavors and photons. We neglect the existence of atomic nuclei and all effects of strong interactions on the equation of state. For a given baryon number density $n_B$, lepton fraction $Y_L$ and specific entropy $S$ we get all other thermodynamic parameters using the condition of beta equilibrium. On Fig. 1 we show the results of our equation of state calculations and compare it with results obtained with the method of Lattimer & Swesty (1991).

## 2.3. Neutrino cooling and annihilation

We use the opacities of Burrows & Lattimer (1986) for neutrinos. To find the neutrino radiation field at a given point we follow back null geodesics in Kerr spacetime until we reach a point in the torus which is at the optical depth $\tau_\nu \approx 1$. This condition is not always met: some rays cross the outer, optically thin parts of the torus, some miss the torus completely. We assume no neutrino radiation is coming from directions represented by such trajectories. We treat separately electron neutrinos, electron antineutrinos and other neutrino flavors, because their opacities and chemical potentials are different. Neutrino opacities and optical depths are proportional to the square of the particle energy, $\tau(E) = \tau_1 E^2$, where $\tau_1$ is calculated for unit energy, for a given neutrino flavor. The Rosseland mean opacity is close to the opacity calculated for the mean square energy and the same is true for the optical depth:

$$\tau_{\text{Rosseland}} \approx \tau_1 <E^2>. \qquad (11)$$

To find the place where a particular ray crosses the neutrinosphere we use the condition $\tau_{\text{Rosseland}} = 1$ integrating $\tau_1$ along the trajectory and substituting the averaged square energy at consecutive points. We assume that neutrinos coming along any ray have the equilibrium distribution given by the temperature $T$ and chemical potential $\mu$ at the neutrinosphere:

$$f_i(E, \mathbf{n}) = \frac{1}{h^3} \frac{1}{\exp\left(\frac{E - \mu_i}{kT_i}\right) + 1} \qquad (12)$$

where h is the Planck constant, k is the Boltzmann constant and $\mathbf{n}$ - the direction of the ray. The subscript $i$ enumerates different kinds of neutrinos and their neutrinospheres can be located at different places, which affects the temperature.

The flux of energy $F$ and angular momentum $H$ caried away by neutrinos and measured by a distant observer can be calculated as:

$$F_i = \int \int cE \, f_i(E, \mathbf{n}) \, d\Omega \, p^2 dp \qquad (13)$$

$$H_i = \int \int clE \, f_i(E, \mathbf{n}) \, d\Omega \, p^2 dp \qquad (14)$$

particle. The integration is over the momentum space and takes into account particles with directions belonging to a small solid angle encompassing the source. Placing observers at different inclination angles $\theta$ and the same distance from the source $r$ we have the luminosity of the source:

$$L_i = 4\pi r^2 \int_0^{\pi/2} F_i(\theta) \sin\theta d\theta. \qquad (15)$$

Replacing $F_i$ by $H_i$ in the above integral one can calculate the angular momentum emission rate $\dot{J}_i$. The total luminosity (neutrino cooling rate) is given as $L_{tot} = L_{\nu_e} + L_{\bar{\nu}_e} + 4L_{\nu_\mu}$. The luminosities of muon and tau neutrinos and antineutrinos are the same because they have vanishing chemical potentials and are produced by the same, neutrally charged reactions. Similarly $\dot{J}_{tot} = \dot{J}_{\nu_e} + \dot{J}_{\bar{\nu}_e} + 4\dot{J}_{\nu_\mu}$.

We find also the neutrino distribution function for several points on the rotation axis. Again, there is no symmetry between electron neutrinos and antineutrinos but the other neutrinos have the same distributions. Following Goodman et al. (1987), Mochkovitch *et al.* (1993) and Jaroszyński (1993) we have the following formula for the energy/momentum deposition rate due to the neutrino annihilation into $e^\pm$ plasma:

$$\dot{Q}_a = 2c \sum_j K_j G_F^2 \times$$
$$\int \int (p_a + \bar{p}_a) f \bar{f} E \bar{E} (1 - \mathbf{n}\bar{\mathbf{n}}) \, d\Omega d\bar{\Omega} p^2 dp \bar{p}^2 d\bar{p} \quad (16)$$

where $\dot{Q}_t$ is the energy deposition rate and $\dot{Q}_r$ is the radial momentum deposition rate. The summation is over neutrino flavors: for electron neutrinos we have $K_e = 0.124$ due to charged and neutral current interactions; for other particles only neutral currents are possible and $K_\mu = K_\tau = 0.027$. $G_F^2$ is the coupling constant for weak interactions. The integration is over the momentum space of neutrinos and antineutrinos, antiparticle variables are denoted by bars. The expression under the integral depends on the flavor but we do not show it explicitly. Except for the $p_a + \bar{p}_a$ factor the expression under the integral is invariant and can be calculated in any frame.

### 2.4. The formation of $e^\pm$ jet

The neutrino annihilation produces $e^\pm$ plasma. The neutrino pairs have typically energies much larger then electron rest mass, so the plasma is relativistic. The momentum is injected into the plasma during the annihilation which also affects its motion. We assume that a $e^\pm$ jet forms in a narrow funnel around rotation axis. For technical convenience we treat the jet as stationary and assume its quasi-spherical symmetry inside a narrow cone. The confinement problem is too complicated, depends on the history of matter outflow from the system and we are not long as the plasma travel time to the sonic point is shorter than the timescale for the evolution of the neutrino source.

Due to our assumptions the jet can move radially out or in. We expect that there is a stagnation point in the flow at some radius $r_0$. Plasma above this point moves outward through the outer sonic point; plasma below undergoes accretion crossing the inner sonic point. The flow is stationary but the fluxes of particles or energy are not constant in space because of the continous and distributed in space injection. The energy and momentum equations have the form:

$$\nabla_b T_a^b = \dot{Q}_a \qquad (17)$$

where $T_b^a = (\epsilon + P) u^a u_b - P \delta_b^a$ is the energy momentum tensor of plasma (e.g. Misner, Thorne & Wheeler 1973) in standard notation and $\dot{Q}_b$ is defined in Eq.(16). For a relativistic pair plasma we have $\epsilon = 3P$ and due to the assumed symmetry of the flow there is one independent component of the fourvelocity. We choose the variable $u$:

$$u^2 \equiv |u^r u_r|; \quad u^t u_t = 1 + u^2 \qquad (18)$$

to define $u^a$. Now the plasma flow in the jet is defined by two independent variables, the pressure $P$ and radial momentum $u$. The Lorentz gamma factor is related to $u$: $\Gamma = \sqrt{1+u^2}$.

The energy and radial momentum equations are sufficient to describe the flow of plasma. We use the Boyer-Lindquist coordinate system in the Kerr spacetime outside the horizon. Only two metric components ($g_{tt}$ and $g_{rr}$) and the metric determinant $g \equiv -A^2 \sin^2\theta$ appear in the equations. In the narrow cone around the rotation axis ($\theta \leq \Delta\theta \ll 1$) we use $g_{tt}$, $g_{rr}$ and $A$ as measured on the axis. Accidentally on the axis one has $g_{tt} g_{rr} = -1$ and we define $g_{rr} \equiv -A/\Delta$ (compare with Bardeen 1973a). The energy equation can now be integrated and after some algebra written in the form:

$$4Pu\Gamma = \frac{L_{\nu\nu}/4\pi - \int_r^\infty \dot{Q}_t A \, dr}{\Delta} \qquad (19)$$

where $L_{\nu\nu}$ is the jet power as measured by an observer on the rotation axis, who assumes the source to be isotropic. (To get the true power in the jet one should multiply $L_{\nu\nu}$ by $\Delta\Omega/4\pi$, where $\Delta\Omega$ is the solid angle subtended by the jet. The radial momentum equation reads:

$$(1+4u^2)P' + 4P\left(2uu' + (1+2u^2)\frac{\Delta'}{\Delta} - \Gamma^2\frac{A'}{A}\right) = \dot{Q}_r \quad (20)$$

where the prime denotes differentiation relative to $r$. The first equation can be used to eliminate the pressure from the second, which becomes a first order differential equation for $u$. It is singular at sonic points, where $u = \pm\sqrt{1/2}$

merator of the RHS of energy equation must vanish which defines the jet power as seen by a distant observer:

$$L_{\nu\nu} = 4\pi \int_{r_0}^{\infty} \dot{Q}_t \, A \, dr \qquad (21)$$

Thus the location of the stagnation point affects the power of the jet as seen from infinity. For a given stagnation point location one can find the positions of sonic points requiring the smooth behaviour of jet velocity (or momentum $u$) there. Next we integrate the momentum equation from sonic points toward the stagnation and require its smoothness at this point as well. Changing the stagnation point and repeating the procedure we find the solution, which is in agreement with all the criteria. The examples of solutions describing the flow in the jet are shown on Fig. 2.

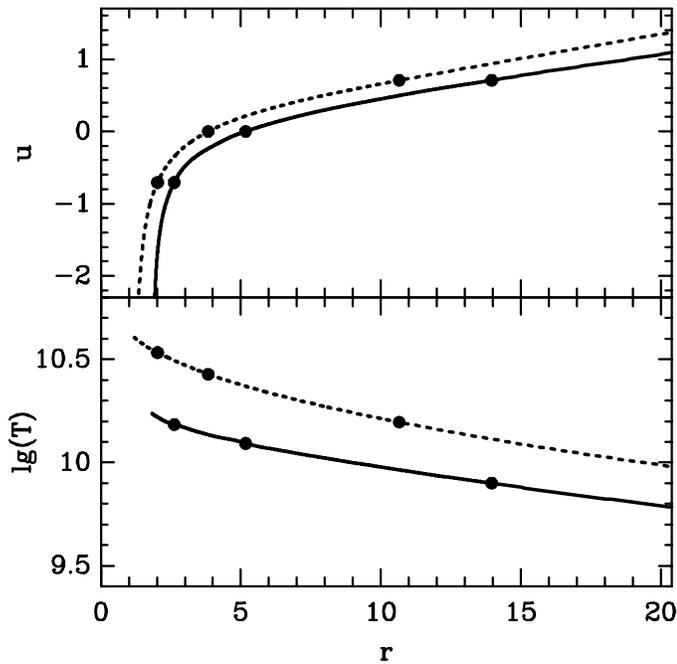

**Fig. 2.** Radial momentum (upper pannel) and the temperature (lower pannel) of the $e^{\pm}$ plasma in jets. Solid lines are for the model "R" with the specific entropy $S = 10$. Doted lines represent model "W2", $S = 10$. (See the next section for the meaning of these symbols). The singular points in the flow are marked on the plots.

## 3. Astrophysically interesting configurations

### 3.1. Tori resulting from mergers

We use some of the results of Rasio & Shapiro (1992, hereafter RS), Davies et al. (1994, hereafter DBPT) and Zhuge et al. (1994, hereafter ZCM) to choose the parameters of tori models. All these calculations of binary neutron stars clusion of gravitational radiation as the only relativistic effect. This approach is dictated by technical limitations – fully relativistic, $3^d$ hydrodynamical calculation of complicated systems is not yet possible. According to the above authors, the merger of neutron stars binary produces a rapidly (but not at breakup speed) rotating central body containing most of the mass and a disk of material with higher specific angular momentum. All the above authors use the $1.4 \, M_{\odot} + 1.4 \, M_{\odot}$ binary as typical in their calculations and the central object resulting from the merger has the mass $\sim 2.2 - 2.5 \, M_{\odot}$. This is above the maximal mass of a rotating neutron star for the majority of equations of state considered by Salgado et al. (1994). Thus it is possible that the central object will subsequently collapse producing a Kerr black hole. The binary system loses angular momentum by gravitational radiation. RS assume that the binary system is rotating synchronously, which means that the spins of neutron stars simply add to the orbital angular momentum. Kochanek (1992) has shown that the viscosity of matter inside neutron stars is too small to keep the synchronization of the system until the merger of the stars. Probably the assumption that the stars have zero spin in the inertial frame is better founded than synchronization. Calculations of DBPT and ZCM use low spin neutron stars. Comparison of the models suggests that the mass of the disk increases with the spin of neutron stars. A more massive disk is needed to accomodate the excess angular momentum. The parameter characterizing angular momentum of the central object (identical with the Kerr $a$ parameter in the case of a black hole) has the value $cJ/GM^2 \approx 0.6$ according to RS, DBPT and ZCM despite some differences in their approaches. This value is typical also for the models of rotating neutron stars with maximal mass and different equations of state calculated by Salgado et al. (1994). This means that their upper limits for rotating neutron star mass are relevant to the situation considered and that the merger can only produce a Kerr black hole with moderate angular momentum.

The amount of energy emitted as gravitational waves calculated by different authors is also in agreement and corresponds to $\sim 0.1 \, M_{\odot} c^2$. The predicted mass of the disk is however different. According to ZCM 6% – 8% of the system mass goes into the disk, depending on the equation of state (size of the single neutron star). DBPT give 5% – 13% in this place (changing the spins of neutron stars from counterrotation to about 1/5 of the value required for synchronization). RS, who assume the binary rotates synchronously, obtain about 20% of the mass in the disk.

We construct three families of models which partially correspond to the mentioned results of merger calculations. The masses of black holes+disks systems are $2.5 \, M_{\odot} + 0.2 \, M_{\odot}$ (sequence "Z") $2.4 \, M_{\odot} + 0.3 \, M_{\odot}$ (sequence "D" and $2.2 \, M_{\odot} + 0.5 \, M_{\odot}$ (sequence "R"). We use

Kerr black hole with the angular momentum parameter $a = 0.6$.

The merger calculations we refer to, give the power law distribution of the disk angular velocity. The value of the power law index given by RS is $\approx 1.8$, DBPT give 1.5 in the disk and 1.8 in its outermost parts (*tails*) while ZCM obtain $\beta \approx 1.7$. We adopt $\beta = 1.8$. It is close to the quoted results and gives the thickest tori and the best chance of obtaining a powerful source of neutrinos.

*3.2. Tori resulting from failed supernova scenario*

According to Woosley (1993) a black hole plus torus system may be the product of the collapse of a massive, rapidly rotating star, a *failed supernova* as he calls it. The evolution of a massive, rotating star depends on the history of mass loss and the amount of angular momentum. Thus the possible parameters of the resulting black hole and disk system are far less constrained as compared to the products of the neutron stars mergers. We consider two families of models "W1" and "W2". In both cases the black hole mass is $M_{BH} = 3 M_\odot$ as suggested by Woosley (1993) and the torus mass $M_{torus} = 1 M_\odot$ - two times more than his estimate. The angular momentum of the hole is difficult to estimate. It is possible that the collapsing core transports part of its angular momentum out, whenever its rotational kinetic energy exceeds 14% of its potential energy (Ostriker & Peebles 1973; RS; DBPT). Such mechanism would produce a black hole with $a \sim 0.6$ as in the case of merging neutron stars and we use this value for W1 sequence of models. For the sequence W2 we assume $a = 0.99$. This high rotation rate may be the result of an axially symmetric collapse with no mechanism of angular momentum transport or the result of consecutive accretion of high angular momentum material. The main reason for introducing high angular momentum of the black hole in the sequence W2 is the investigation of the dependence of our results on this parameter.

## 4. Results

*4.1. Initial configurations*

First we investigate the efficiency of producing the relativistic plasma jets by our configurations. The critical parameter is the temperature at the neutrinosphere since it defines the energy and spectrum of neutrino flux. In our approach we use the entropy per baryon as a parameter to define the equation of state. We construct sequences of models with increasing specific entropy $S$ for families of tori described in the previous section. Models of each sequence have the same baryon masses in the disk and the same black hole masses. All tori are placed as close to the black hole as possible, so their inner edges are on Keplerian orbits between marginally bound and marginally stable radii (AJS). The volumes of the tori increase with slightly closer to the black hole and the outer radii become greater. For models with very high specific entropy, $S > 12$, the neutrinosphere becomes smaller with increasing entropy. This is because higher entropy implies lower density of matter and lowers the optical depth. The angular momentum parameter is kept constant in each sequence: $\beta = 1.8$. The total angular momentum increases with $S$ as a consequence of wider matter distribution. The efficiency is defined as:

$$\text{eff} = \frac{L_{\nu\nu}}{L_{tot}} \qquad (22)$$

- the ratio of the jet power (rescaled to $4\pi$) to the rate of the neutrino cooling. The efficiency as a function of the specific entropy is shown in the upper panel of Fig.3 for various families of tori. As one can see, a relatively high efficiency of converting the disk energy into jet energy is possible for models with high specific entropy, approximately in the range $8 < S < 12$. Outside this interval the efficiency is sharply reduced. For low entropy models the temperature at the neutrinosphere becomes small, which reduces the neutrino luminosity and annihilation rate. Very high entropy models have low density and become optically thin.

The efficiency of our models decreases sharply with decreasing specific entropy. If viscous mechanisms are absent or inefficient one can expect the initial configurations to cool via neutrino emission on the scale $\tau_{cool} = E_{torus}^{int}/L_{tot}$. The energy taken to infinity by the jet in such a case can be estimated as:

$$E_{\nu\nu} = \tau_{cool} \, L_{\nu\nu} \, , \qquad (23)$$

which assumes the initial efficiency to remain constant. In the middle and lower panels of Fig.3 we plot the initial power of jets $L_{\nu\nu}$ and the estimated energy taken to infinity from a cooling torus $E_{\nu\nu}$ as functions of the initial specific entropy for various model sequences.

*4.2. Evolutionary sequences*

If some viscosity mechanism exists, the tori may evolve remaining hot. This requires the balance between heating and cooling, $Q_{visc}^+ \approx L_{tot}$ and may be used to estimate the viscosity parameter $\alpha$. The friction causes changes in angular momentum distribution. For configurations which have the inner edges on unstable Keplerian orbits, part of the mass falls toward the horizon (AJS). Using our simplified configurations with only one parameter defining the angular momentum distribution ($\beta$) and one defining the thermal properties of matter ($S$) we are able to construct quasi-evolutionary sequences of models which change under the action of viscosity. We have to terminate our calculations before the whole torus mass is accreted, because for low values of parameter $\beta$ ($\leq 1.72$ in the case of slowly rotating black holes with $a = 0.6$), the inner edge of the disk cannot be placed on an unstable Keplerian orbit. We consider sequences of efficient configurations with the high value of specific entropy $S = 8$ or 10. We also calculate a sequence with higher initial value of the parameter $\beta = 1.9$, to see the influence of the initial angular momentum distribution in the disk on its ability to produce a powerful fireball.

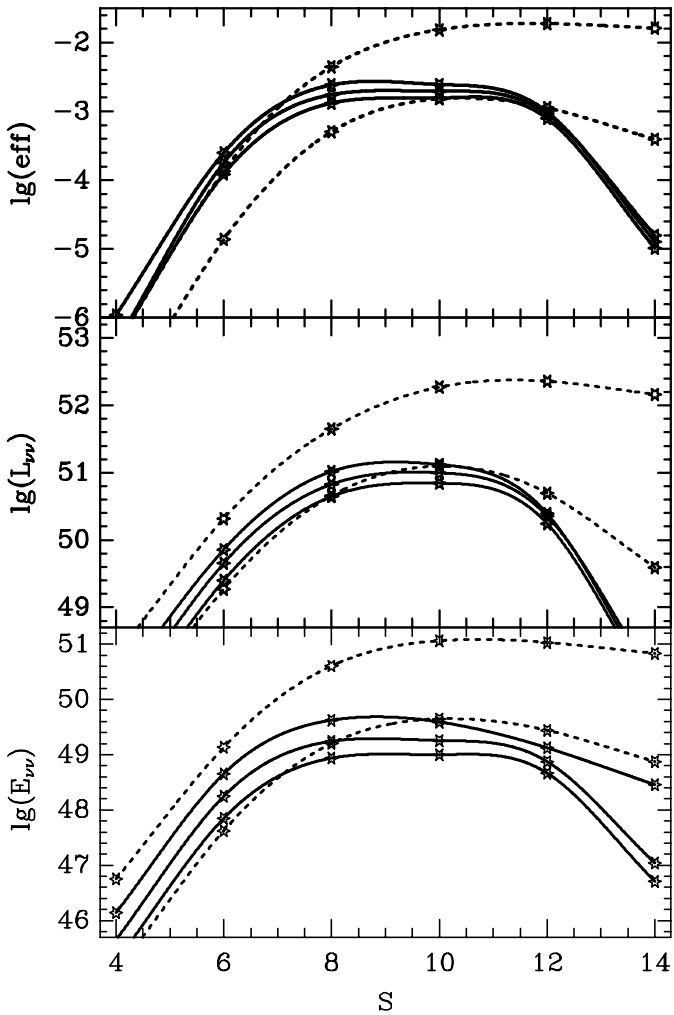

**Fig. 3.** Efficiency (upper panel), jet power (middle) and energy transported to infinity by the jet (lower panel) as functions of the initial specific entropy of the models. Solid lines are used for the merger models "Z", "D", "R" (bottom to top) and dotted lines for the *failed supernova* models "W1" and "W2" in each panel. The lines are drawn with the help of the cubic spline interpolation through the calculated points which are marked.

The global parameters defining the tori along the quasi-evolutionary sequence are calculated from the conservation laws. Suppose a small part of torus baryon mass $\Delta M$ is accreted onto the black hole from the inner edge (Thorne 1974):

$$\Delta M_{BH} = \Delta M\, u_{in} \qquad (24)$$

$$\Delta J_{BH} = \frac{GM_{BH}\Delta M}{c}\, l_{in} u_{in} \qquad (25)$$

where $u_{in} \equiv |u_t(r_{in})|$ and $l_{in} u_{in} \equiv |u_\phi(r_{in})|$ are respectively the energy and angular momentum of a unit mass particle orbiting on the inner edge. We neglect the energy and angular momentum transport into the hole by neutrinos.

The region opaque to neutrinos is concentrated around the Keplerian orbit at $r_c$, where the pressure is maximal. We estimate the specific angular momentum caried away by neutrinos to be $l_c$ - this is a good guess, which can be checked by a direct calculation of $\dot{J}_{tot}$ and $L_{tot}$. When the amount $\Delta M$ of the rest mass is accreted, approximately $\Delta M(1 - u_{in})$ is radiated away from the system and this part takes the angular momentum to infinity. The change in the torus angular momentum can be described as:

$$\Delta J_{torus} = -\frac{GM_{BH}\Delta M}{c}\left(l_{in} u_{in} + l_c(1 - u_{in})\right). \qquad (26)$$

Two terms represent the angular momentum caried to the black hole by the matter and radiated away with neutrinos. Iteratively we find new torus model with the required mass and angular momentum. Specific entropy is kept constant by assumption. The new value of $\beta$ results from iterations.

For each model in the sequence we can calculate the total energy of the system:

$$E_{sys} = M_{BH}c^2 + E_{torus} \qquad (27)$$

where we neglect the perturbation to the black hole caused by the torus (cf Bardeen 1973b). The time for each evolutionary step can be estimated as $\Delta t = \Delta E_{sys}/L_{tot}$. Using this relation we can get the time dependence of any quantity characterizing an evolving torus model.

In Fig.4 we show the dependence of internal model parameters on time. The displayed parameters are: the angular momentum parameter $\beta$, the required value of viscosity parameter $\alpha$ and the black hole angular momentum parameter $a$. The Kerr parameter of the black hole increases due to the accretion. For initial configurations with $a = 0.6$ it remains moderate ($a < 0.8$). Since effectiveness of our models is better for high angular momentum black holes (see Fig.3) increasing $a$ increases the power delivered to the jet. The main mechanism of losing angular momentum by the torus is the accretion of low specific angular momentum matter from the inner edge. The averaged specific angular momentum of the remaining part increases and the parameter $\beta$ decreases. (Thus the matter is transported inward and angular momentum outward - compare Lynden-Bell & Pringle 1974). The required value of $\alpha$ depends in complicated way on the structure of models. It has to be high ($\alpha \geq 0.1$ or even $\alpha \geq 0.3$ in some cases) if the angular momentum transport and accretion is to play any role in powering the jets.

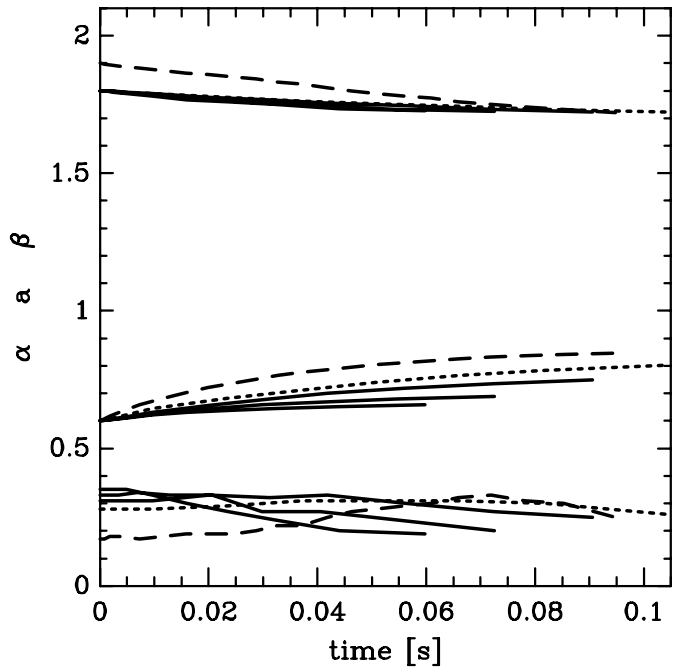

**Fig. 4.** The values of the angular momentum distribution parameter $\beta$, Black hole angular momentum parameter $a$ and the required viscosity parameter $\alpha$ shown as functions of time for chosen evolutionary sequences. Solid lines are for the sequences starting from merger configurations "R", "D" and "Z" with the initial entropy $S = 10$. The dotted line shows the results for the configuration "W1". The dashed line illustrates the evolution of a configuration, which has a different initial angular momentum distribution with $\beta = 1.9$ and all other parameters of configuration "R" with $S = 10$. (See the text for further details).

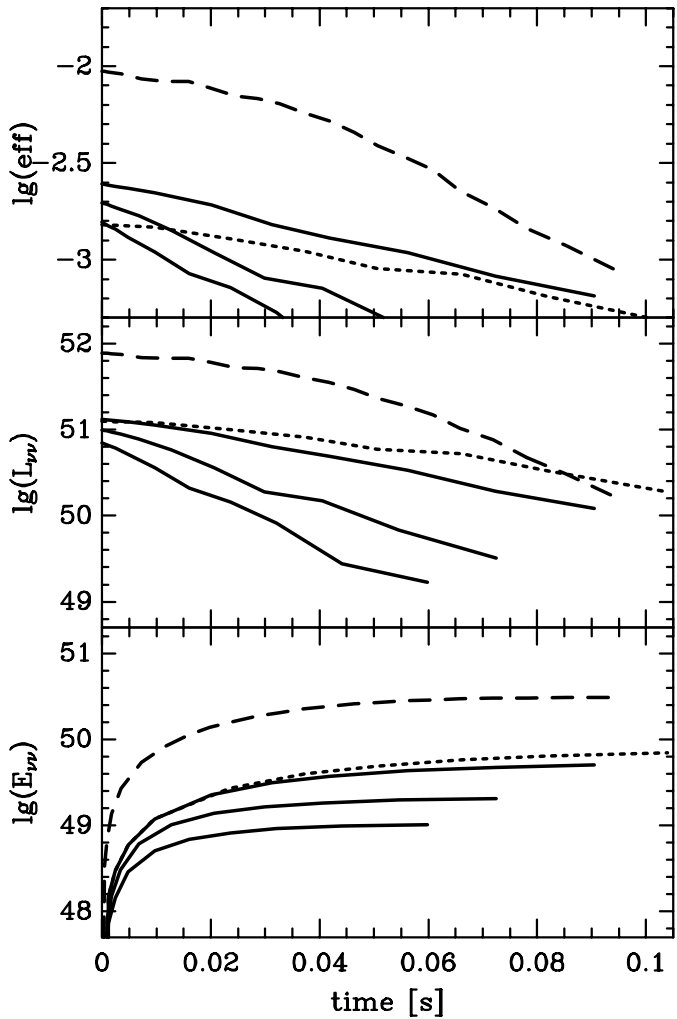

**Fig. 5.** Efficiency (upper panel), jet power (middle) and energy transported to infinity by the jet (lower panel) as functions of time for the chosen evolutionary sequences. The drawing conventions are the same as on Fig.4 and further details are in the text.

Following the quasi-evolutionary sequences we see that the efficiency of models decreases with time. Also the power of jets $L_{\nu\nu}(t)$ is a decreasing function of time. Integrating the power over time we get the cumulative energies radiated by the jets since the begining of the burst $E_{\nu\nu}(t)$. Examples of such dependences are shown on Fig.5. The configuration starting from a "R" model with $\beta = 1.8$ evolves to $\beta \approx 1.72$ (where we have to stop our calculations) in $\approx 0.09$ s emitting $5 \times 10^{49}$ ergs through the jets and losing 1/3 of the initial mass. The efficiency drops by a factor 4 during this time, so extrapolation would give the total energy emitted via the jets (rescaled to $4\pi$) $E_{\nu\nu} < 10^{50}$ ergs. For configurations starting from "D" or "Z" models the drop in efficiency is even faster and we get $E_{\nu\nu} < 3 \times 10^{49}$ ergs and $E_{\nu\nu} < 1 \times 10^{49}$ ergs respectively. For a sequence starting from the model "W1" with $S = 10$ we get $E_{\nu\nu} < 10^{50}$ ergs. We are not calculating a sequence starting from a "W2" model, because such models can produce $E_{\nu\nu} \approx 10^{51}$ ergs by cooling alone (cf. Fig.3) and it is not necessary to look for effects that would make this quantity higher. The model which has $\beta = 1.9$ initially is shown for comparison. We can follow its evolution up to a point, where 3/4 of the torus mass is accreted. It has relatively high efficiency and delivers $E_{\nu\nu} \approx 3 \times 10^{50}$ ergs to the jets. We have also calculated evolutionary sequences of models with $S = 8$ for some configurations. The results are very similar to the $S = 10$ case.

## 5. Discussion and conclusions

The model we consider in this paper is highly simplified. We assume that the central body being a product of the

Kerr black hole. This is not necessarily true for all possible equations of state; in particular our models based on RS paper with the black hole mass $M_{BH} = 2.2\ M_\odot$ may be irrelevant. For other configurations ($M_{BH} \geq 2.4\ M_\odot$) this assumption is less restrictive.

The distribution of the specific angular momentum, that we use for our tori, has only one free parameter. In Newtonian limit it corresponds to the power law distributions of angular momentum and angular velocity with radius. Similar shapes of angular momentum distribution in the disks, being the result of neutron star mergers are obtained by RS, DBPT and ZCM.

Sequences of our models with changing angular distribution parameter $\beta$ can be used to describe the evolution of tori under the action of viscosity. Lower $\beta$ means flatter configurations, $\beta = 2$ corresponds to a constant specific angular momentum torus. Our scheme allows for the description of viscous tori evolution, when it is accompanied by accretion. In such a case the inner edge of the torus must remain on an unstable Keplerian orbit (AJS), which limits the possible values of angular momentum distribution parameter to $\beta \geq 1.72$. This means that we are not able to follow the disk evolution all the way up to a Keplerian, flat configuration. On the other hand our calculations show, that the disks are less effective as neutrino emmiters, when they become flat, so the late stages of the evolution are not interesting for the modeling od gamma ray bursts.

We investigate the simplified, isoentropic configurations, changing the specific entropy per baryon in the range $4 \leq S \leq 14$. Jaroszyński (1993) and Witt et al. (1994) use $S \leq 3$ but their models are too cold, to convert a significant part of the energy into a relativistic jet. The maximal temperature in the torus obtained by RS is $\sim 10$ MeV, DBPT get much cooler configurations. It is probably necessary to heat up the torus by viscosity, after its formation. When the torus gets hot the convection may play some role in transporting neutrinos as proposed by Herant et al. (1994) for supernovae. This mechanism can also work toward an isoentropic configuration.

We check also the influence of entropy distribution within the configuration on our results. The method of constructing equlibrium models, which we use, works only for barotropic equations of state (AJS). We investigate several equations of state with specific entropy being a given function of the particle density. Such configurations are barotropic. The models with the specific entropy of matter decreasing with the density are convectively stable; when the specific entropy increases with the density, the opposite is true. We use both kinds of models since our method of calculating neutrino luminosity and annihilation rate works regardless of the configuration stability. Our calculations show, that introducing a stable entropy gradient into the configuration lowers the efficiency of producing the jet, and the unstable entropy gradient increases gated configurations with the specific entropy changing no more than by the factor 2), the isentropic configurations are the most efficient among stable ones.

We do not consider the energy transport inside the configuration. We use only the condition of a global cooling/heating balance assuming the specific entropy to remain constant. Neutrino cooling from the neutrinosphere ($\tau_\nu \approx 1$) must be balanced by viscous heat production inside the torus. The neutrino creation by various processes inside the torus (Itoh et al. 1990) is fast enough to keep their population inside the neutrinosphere in equilibrium despite the fast cooling rate.

The problem of jet confinement is beyond the scope of this paper. We assume that a jet can form in a conical region around the rotation axis, where baryons are absent. The size and the shape of this region should be better constrained by investigation of the wind from the torus surface and its interaction with the jet. (Problem of wind formation near the surface of a hot neutron star is described by Salpeter & Shapiro 1981 and by Duncan et al. 1986. The axial symmetry of our problem makes the wind modeling much more difficult.) The shape (opening size) of the jet has no direct influence on the characteristic of radiation coming to an observer near the rotation axis; the probability of observing the source by a randomly placed observer is however directly proportional to the solid angle covered by the jet. Jaroszyński (1993) investigates the neutrino annihilation not only on the rotation axis but also in its surounding and finds no strong dependence of energy deposition rate on position. Thus the wider jet has a better observability remaining intrinsically similar.

The initial configurations we construct are based on results of hydrodynamical, Newtonian calculations by RS, DBPT and ZCM. We follow their results in choosing the masses for the torus and mass and angular momentum for the blackhole. The initial value of $\beta = 1.8$ is the maximal value quoted. Changing the specific entropy of our models we are forced to change their volume, outer radius and the total angular momentum. Low density regions of hot tori extend to distances few times larger than models of RS or DBPT. The alternative way to construct our models would be to change the angular momentum of the black hole in a way keeping the total angular momentum in each of "R", "D" or "Z" sequences constant. This would, however, couple the influence of the Kerr parameter and specific entropy on the results, which we prefer to avoid. In models related to *failed supernowa* scenario the parameters are much less constrained and we choose two representative values of the Kerr parameter and the torus mass twice larger than Woosley (1993) suggestion.

We have investigated the dependence of relativistic jet power $L_{\nu\nu}$ on the specific entropy of the model. ($L_{\nu\nu}$ would be the jet power, if it covered the full solid angle. The true power is smaller in proportion to the solid angle covered by the jet.) We see that the characteristic in origin, $\sim 10^{51}$ ergs s$^{-1}$ can be reached by our models if they are made hot enough. Simultanously the models cool emitting neutrinos with luminosity $L_{\rm tot}$. If viscosity is unimportant, the whole internal energy is radiated in few $\times 10^{-2}$ s. The total energy transported by the jet is $E_{\nu\bar\nu} \sim$ few $\times 10^{49}$ ergs for models surrounding moderate angular momentum black holes ($a = 0.6$). Only the model around a fast rotating hole ($a = 0.99$) is effective enough to emit $E_{\nu\bar\nu} \sim 10^{51}$ ergs - amount sufficient for a burst model.

We have followed the evolution of our models under the action of viscosity employing a simplified approach. We have checked the possibility of emitting larger amounts of energy through the jets if the tori are continuously heated by dissipation of rotational energy. We have found that the accretion and the transport of angular momentum out make the tori less massive, flatter and more distant from the black hole on average. All these effects decrease the effectiveness of producing the jets. The only effect in the opposite direction is the increase of the black hole angular momentum by the accreted mass. We are not able to follow the evolution till all the mass is accreted (see above), but the extrapolation of the results shows, that most of the energy is already delivered to the jets when we terminate calculations. Our evolutionary paths cover about 100 ms. This can be compared with the time which passes from the first contact of the neutron stars in a coalescing binary to the formation of the disk. In DBPT calculations it takes $\sim$ 3.7 ms and we are not expecting large amount of energy to be emitted during this time. Our final results are shown on Fig.5: the energy emitted by the jets from configurations with moderate angular momentum black holes is still low, smaller than $E_{\nu\bar\nu} \approx 10^{50}$ ergs.

To see the dependence of our results on various model parameters, we investigate a model which initially has $\beta = 1.9$ and other parameters of the "R" configuration. As can be seen on Fig.5 this model gives $\sim 7$ times more effective energy to the jet as compared with its $\beta = 1.8$ counterpart. Clearly models closer to a constant specific angular momentum configurations ($\beta = 2$) are more effective. They have also a greater content of kinetic energy, which can be dissipated. We should stress here, however, that all merger calculations give $\beta \leq 1.8$.

The effectiveness increases also as a result of increasing black hole angular momentum. This is probably a geometrical effect - with higher $a$ the torus can be on average closer to the rotation axis and a larger fraction of neutrinos can annihilate. This also implies that one should use relativistic calculations, when modeling the tori. In Newtonian calculations of neutron star mergers (RS, DBPT and ZCM) one obtains a moderately rotating central object, corresponding to $a = 0.6$. It is also known (Chandrasekhar 1970), that axisymmetric bodies with too much rotation energy ($T/|W| > 0.14$) are unstable and emit gravitational waves, losing angular momentum. During the col-

*supernova* this process is probably not fast enough and a black hole with $a \sim 1$ is not excluded.

In conclusion we see, that the existing merger calculations assuming that a binary consists of small spin neutron stars (DBPT and ZCM) produce tori, which are not likely to be the sources of gamma ray bursts, if they are cosmological. This conclusion has been reached by DBPT, who say that their tori are too cold to effectively emit neutrinos. We go further in this conclusion saying that even if a mechanism of heating up the tori and a mechanism providing the angular momentum transport were known, the energy delivered to the relativistic jet would be an order of magnitude too low to explain bursts. The merger calculation assuming the binary to be rotating synchronously has no foundation (Kochanek 1992), so the "R" model should not be used.

The *failed supernova* scenario (Woosley 1993) is less restrictive. If it can produce a system consisting of a rapidly rotating black hole ($a \sim 0.99$) with a massive torus around it - our calculations do not contradict the possibility of delivering the right amount of energy to the jet. That such configurations can really be produced during a collapse of a WR star remains to be shown in a hydrodynamical calculation.

Also further calculations of the neutron star mergers, including more thermodynamic details and possibly relativistic effects are needed. It is still possible that the disk produced in such event has different parameters than obtained by existing calculations.

*Acknowledgements.* The author wishes to thank Ramesh Narayan, Robert Mochkovitch and Bohdan Paczyński for the discussions. This work was supported in part by Centre National de la Recherche Scientifique through the grant PICS "Astronomie Pologne" No 198 and also by the Polish State Committee for Scientific Research grant 2-P304-006-06.

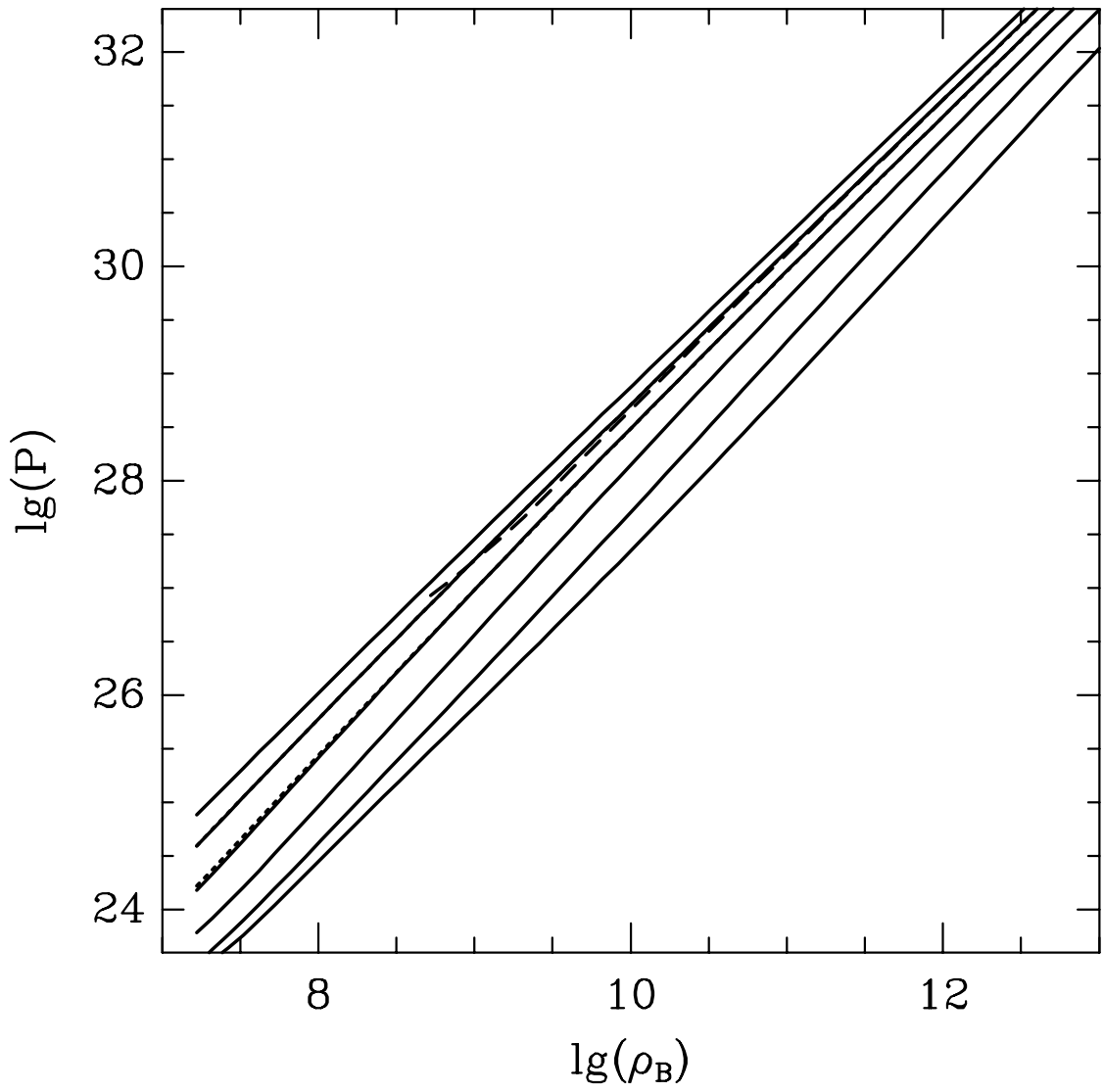

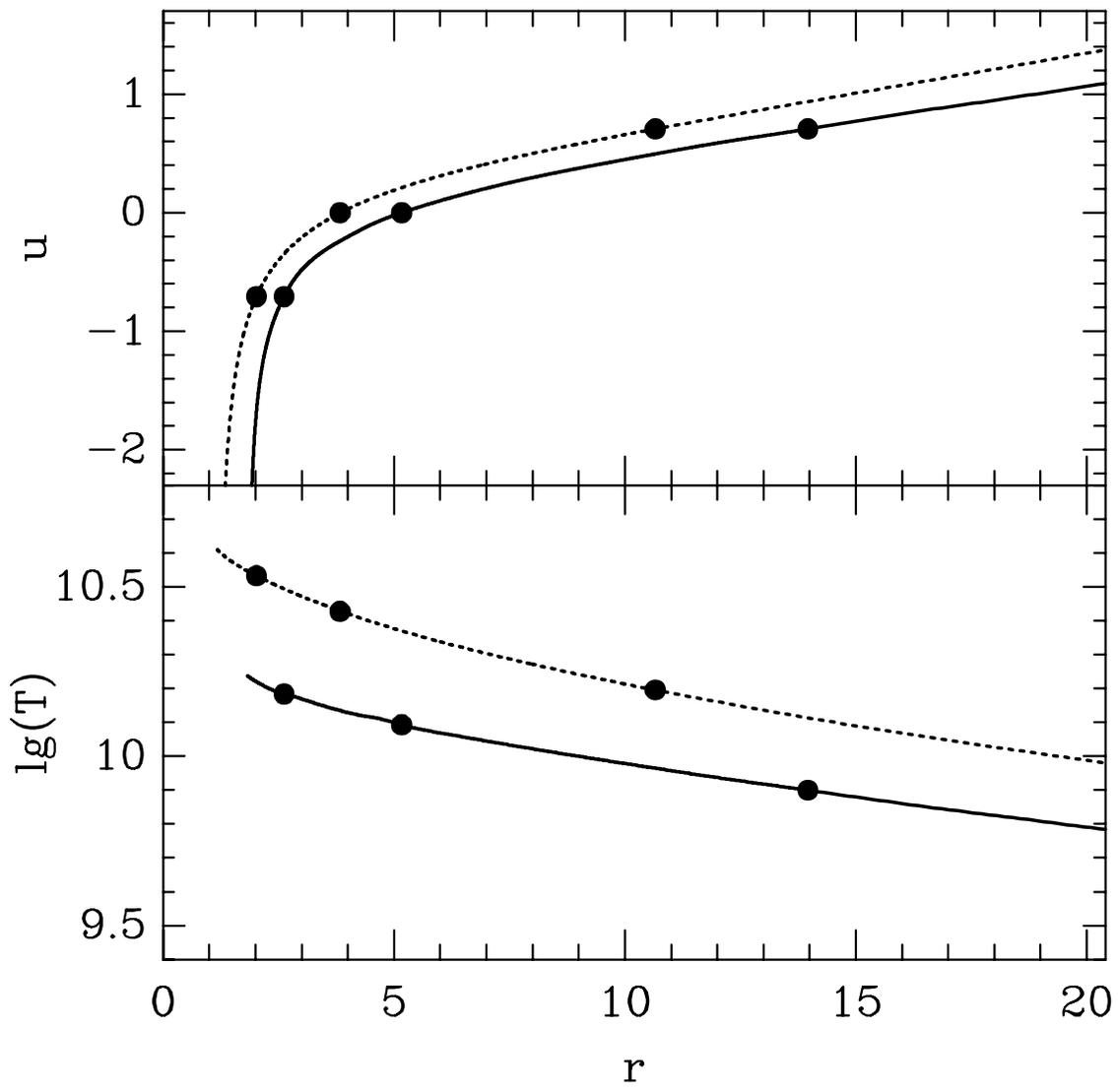

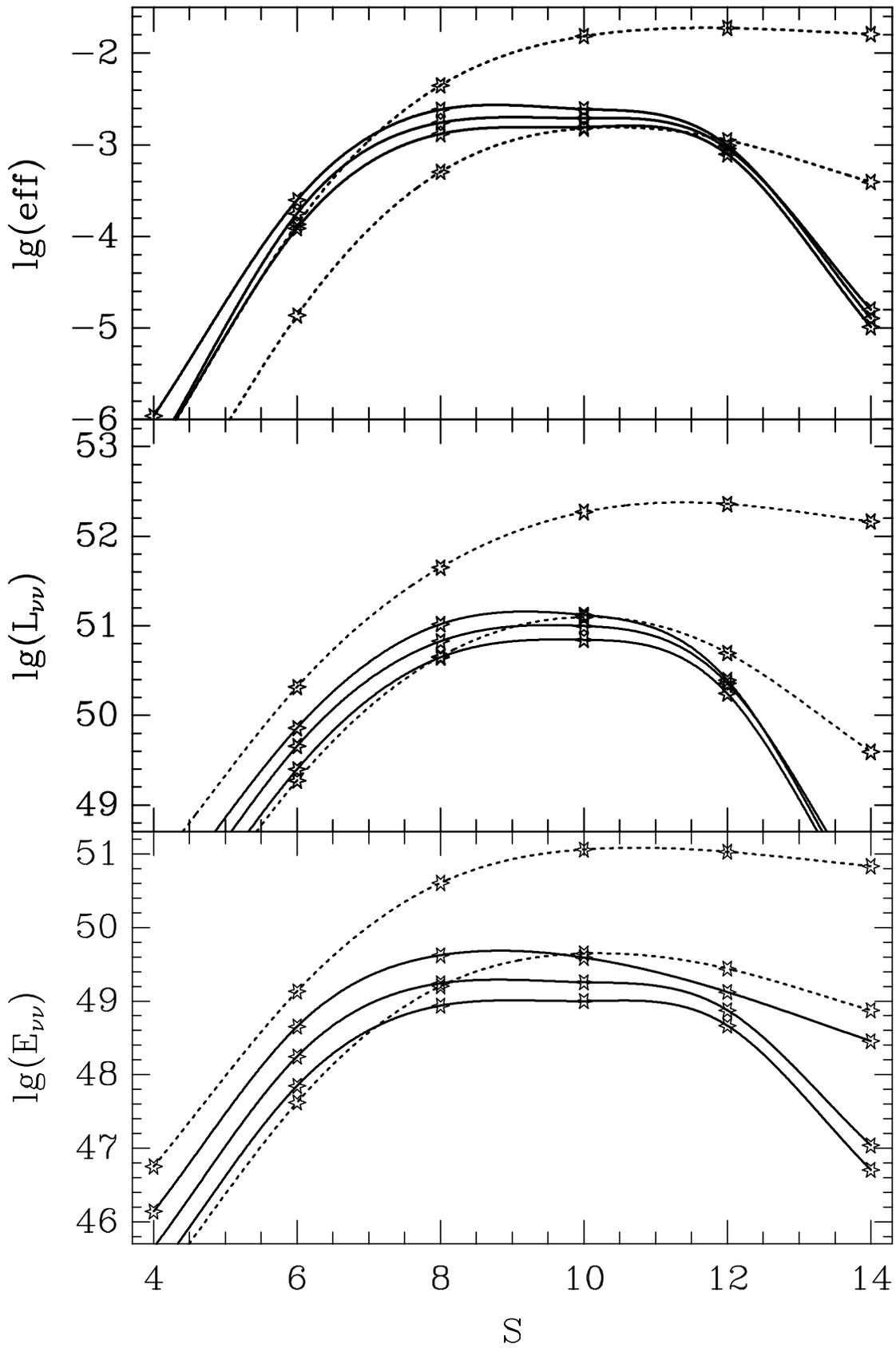

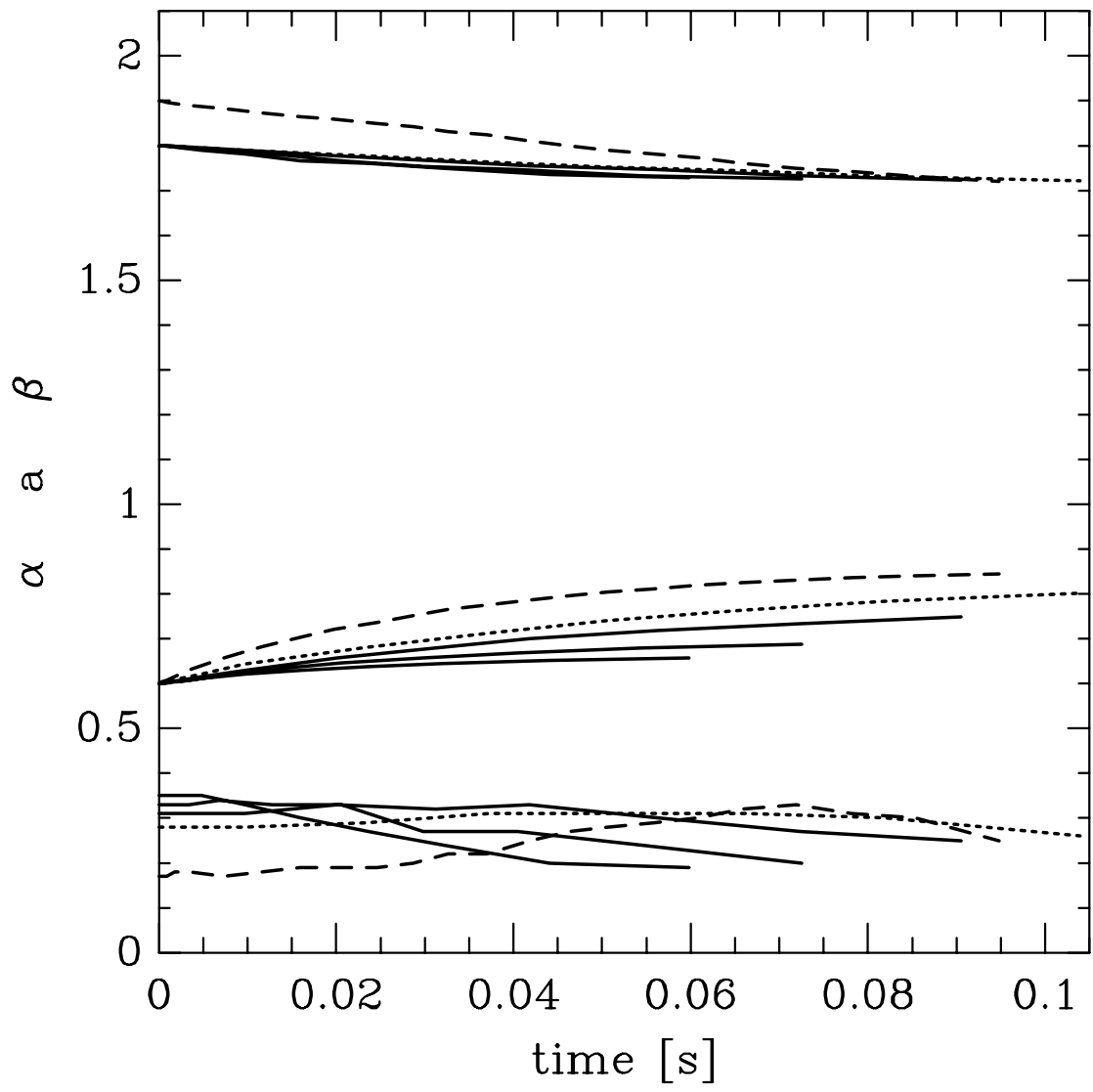

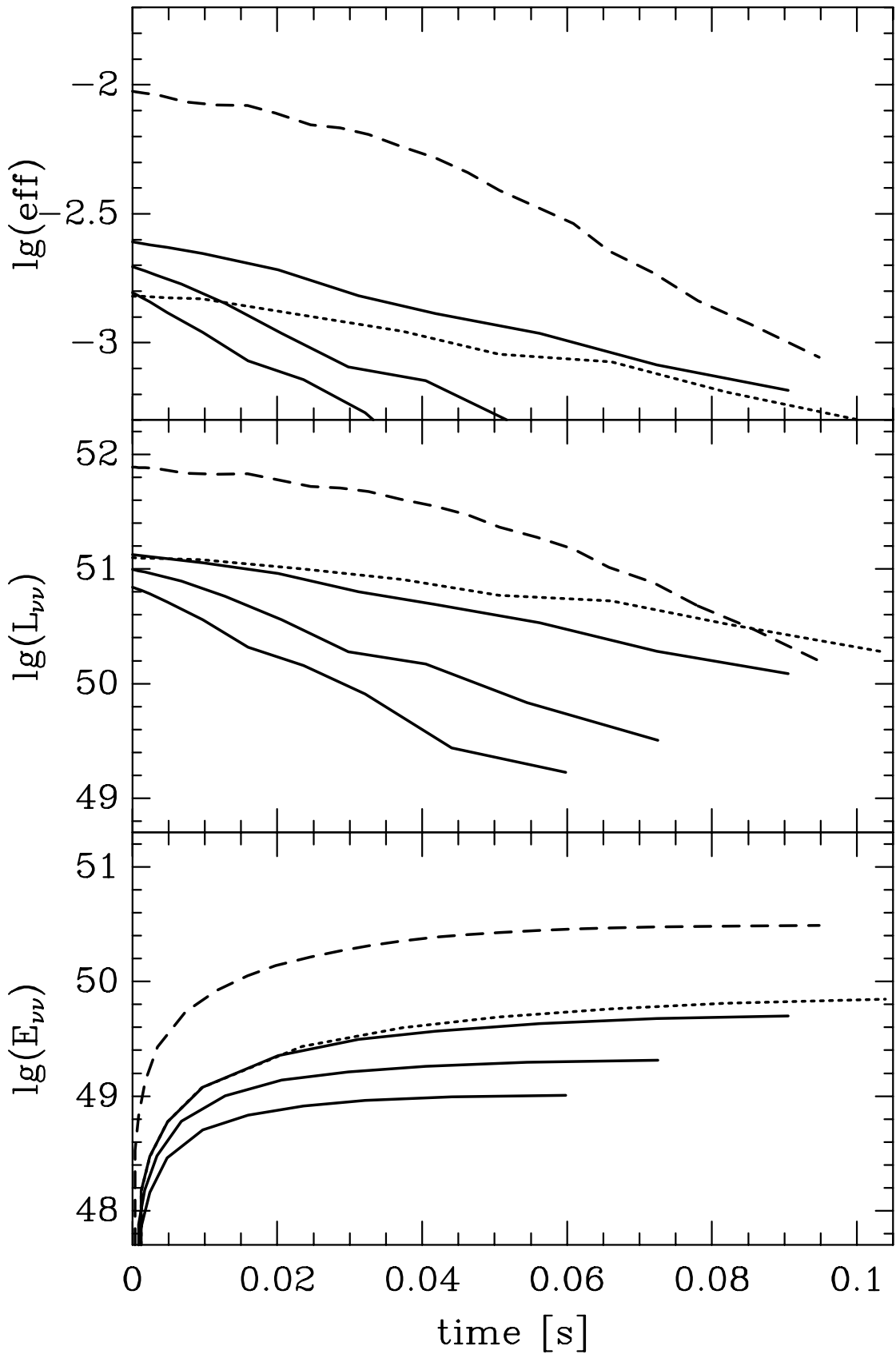